# DEEP Q-LEARING-DRIVEN POWER CONTROL FOR ENHANCED NOMA USER PERFORMANCE

Bach Hung Luu[1], Sinh Cong Lam[1], Nam Hoang Nguyen[2]

[1] Faculty of Telecommunication System, VNU University of Engineering and Technology,
[2] Faculty of Information Technology, East Asia University of Technology

## ABSTRACT

*Cell-edge users (CEUs) in cellular networks typically suffer from poor channel conditions due to long distances from serving base stations and physical obstructions, resulting in much lower data rates compared to cell-center users (CCUs). This paper proposes an Unmanned Aerial Vehicles (UAV)-assisted cellular network with intelligent power control to address the performance gap between CEUs and CCUs. Unlike conventional approaches that either deploy UAVs for all users or use no UAV assistance, our model uses a distance-based criterion where only users beyond a reference distance receive UAV relay assistance. Each UAV operates as an amplify-and-forward relay, enabling assisted users to receive signals from both the base station and the UAV simultaneously, thereby achieving diversity gain. To optimize transmission power allocation across base stations, we employ a Deep Q-Network (DQN) learning framework that learns power control policies without requiring accurate channel models. Simulation results show that the proposed approach achieves a peak average rate of 2.28 bps/Hz at the optimal reference distance of 400m, which represents a 3.6% improvement compared to networks without UAV assistance and 0.9% improvement compared to networks where all users receive UAV support. The results also reveal that UAV altitude and reference distance are critical factors affecting system performance, with lower altitudes providing better performance.*

## 1. INTRODUCTION

In recent years, wireless communication technologies have experienced remarkable development. According to recent studies [1], global mobile data traffic is projected to increase exponentially toward 2030 due to the popularity of smartphones, Internet-of-Things (IoT) devices, and data applications such as video streaming and real-time services. This explosive growth has encouraged continuous development in network architecture, resource allocation, and physical layer design.

However, the deployment of cellular systems in complex geographical environments introduces new challenges. Specifically, the complexity of propagation environments leads to significant performance differences among users [2], [3]. Particularly, Cell-Center Users (CCUs) who are near or have Light-of-Sights to the serving BS and then experience favorable channel conditions can achieve acceptable performance. In contrast, Cell-Edge Users (CEUs) who are positioned at far distances from the BS or cover physical obstructions such as building, human, furniture, and then suffer from significant signal degradation, which consequently results in substantially low performance metrics [4]. Moreover, the performance imbalance between CCUs and CEUs cannot be ignored when Base Stations (BSs) serve the complex environments with heterogeneous obstacle distributions. Therefore, the problem of balancing the performance between CCUs and CEUs should be carefully investigated, and a large number of research works have been conducted to address this challenge.

To address these limitations, Unmanned Aerial Vehicles (UAV) - assisted cellular networks have been introduced as a promising solution. In these systems, UAVs act as aerial relay nodes to support BSs in serving users, particularly CEUs [5]. Unlike BSs, UAVs can be quickly deployed at suitable altitudes and locations to provide high-quality air-to-ground communication links, and then extend coverage to isolated regions. The high positioning of UAVs reduces path loss and shadowing effects, while their flexible mobility allows dynamic adjustment to changing traffic patterns and network conditions. Moreover, UAVs can be moved to optimal positions based on user distribution and channel conditions, which cannot be achieved with fixed regular BSs. Therefore, UAV-assisted communication has attracted considerable research attention, with applications in emergency response [6], urban connectivity [7], temporary capacity support [8], and cell-edge performance improvement [9].

## 2. RELATED WORKS

The application of UAVs in wireless communications has attracted considerable research attention over the past decade. Early investigations focused on optimal UAV placement to maximize coverage or minimize transmission power. The problem of determining the optimal UAV altitude to maximize the number of covered users was studied, demonstrating that an optimal altitude exists due to the trade-off between larger coverage area and path loss [10]. Building upon this foundation, the framework was extended to multiple UAVs, and optimization algorithms for joint altitude and position design were proposed [8]. While these placement strategies achieve notable coverage improvements, they assume static scenarios and do not address dynamic user mobility or interference management in multi-cell environments.

Trajectory optimization for mobile UAVs has been extensively investigated to improve coverage quality over time. Energy-efficient UAV trajectory design that minimizes mission completion time while satisfying throughput requirements was proposed [11]. Joint trajectory and communication design using successive convex approximation was then developed to further enhance system performance [12]. Although trajectory optimization methods demonstrate significant performance gains, they require centralized computation with complete channel state information and accurate knowledge of user locations. These assumptions are difficult to satisfy in large-scale practical deployments. In addition, several studies have investigated UAVs functioning as relay nodes to assist ground communications. Optimal relay positioning for UAV-assisted point-to-point links was analyzed, and closed-form expressions for altitude and location optimization were derived [13]. UAV-assisted cellular networks where aerial platforms serve as flying base stations to offload traffic from terrestrial infrastructure were then proposed [14]. While addressing multi-user scenarios, these approaches assume UAVs operate as independent access points rather than cooperative relays. Thus, they fail to exploit diversity combining benefits when users simultaneously receive signals from both terrestrial and aerial links.

Power control represents a critical mechanism for interference management in cellular networks. Traditional approaches include game-theoretic methods and non-convex optimization techniques [15,16]. These methods typically require complete channel state information and have limited scalability to large networks. Recent advances in machine learning have opened new directions for intelligent resource allocation. Deep reinforcement learning has been applied to spectrum sharing and power control in heterogeneous networks [17,18]. In [19], DQN was employed for energy-efficient resource allocation in V2V communications, achieving better energy performance compared to heuristic methods. In [20], a DQN-based approach was utilized for uplink power control in heterogeneous 5G networks, which significantly improved both QoS and energy efficiency. In [21], a Double DQN-based channel assignment scheme was proposed to enhance spectrum sharing efficiency in densely deployed Wi-Fi/LTE heterogeneous networks, achieving substantial improvements in average throughput. The authors in [22] proposed a distributed multi-agent DRL framework for transmit power control, outperforming conventional centralized schemes. For UAV-assisted scenarios specifically, reinforcement learning was employed to jointly optimize UAV trajectory and power allocation [23]. However, this approach assumes uniform UAV assistance for all users, which may lead to inefficient resource utilization when certain users do not require aerial support.

Despite these important contributions, several fundamental challenges remain unresolved in the existing literature. A main limitation of current approaches is the indiscriminate deployment of UAV assistance across all users within the coverage area. This strategy is inefficient because CCUs already experience strong communication links and possibly gain minimal benefit from additional communication assistance, while the corresponding UAV transmissions introduce unnecessary interference to neighboring cells. Furthermore, most existing relay frameworks enforce binary user association, where users connect either to the terrestrial BS or to the UAV exclusively. This neglects the substantial diversity gain achievable through simultaneous dual connectivity. Another critical challenge concerns the scalability of optimization methods to large networks with numerous BSs and UAVs. In these scenarios, traditional convex optimization and game-theoretic approaches suffer from high computational complexity and require accurate channel models. Finally, the unique characteristics of aerial-terrestrial hybrid networks, including altitude-dependent path loss, and coupled ground-to-ground and air-to-ground interference, require new optimization frameworks that existing terrestrial-only methods cannot adequately address.

In this paper, we address these challenges through a UAV-assisted cellular network architecture with intelligent power control. Our approach employs a distance-based criterion wherein only CEUs - those beyond a threshold distance from their serving BS - receive UAV relay assistance, while CCUs continue to be served by terrestrial BSs only. Each UAV operates as an amplify-and-forward relay, enabling assisted users to simultaneously receive signals from both the BS and the UAV, and then achieve spatial diversity gain. To optimize transmission power allocation across BSs under interference constraints, we employ a Deep Q-Network (DQN) learning framework that learns near-optimal policies without requiring accurate channel models. Simulation results demonstrate substantial performance improvements for CEUs while maintaining service quality for CCUs.

The rest of this paper is organized as follows. Section 3 presents the system model, including network topology, user classification, UAV deployment strategy, channel model, and problem formulation. Section 4 describes the Deep Q-Learning framework, including state-action space design, reward function, and training algorithm. Section 5 presents simulation results and performance analysis. Finally, Section 6 concludes the paper and discusses future research directions.

## 3. SYSTEM MODEL

We consider a downlink cellular network consisting of hexagonal cells deployed in a regular cellular network topology, where each cell has a radius of $R$ as illustrated in Figure 1. At the center of each cell, a BS is equipped with omnidirectional antennas to provide coverage for its associated mobile users. The network operates under a frequency reuse factor of one, wherein all BSs transmit simultaneously over the same frequency band. While this frequency reuse-1 scheme maximizes spectral efficiency, it introduces significant Inter-Cell Interference (ICI), which is a fundamental performance-limiting factor in such deployments.

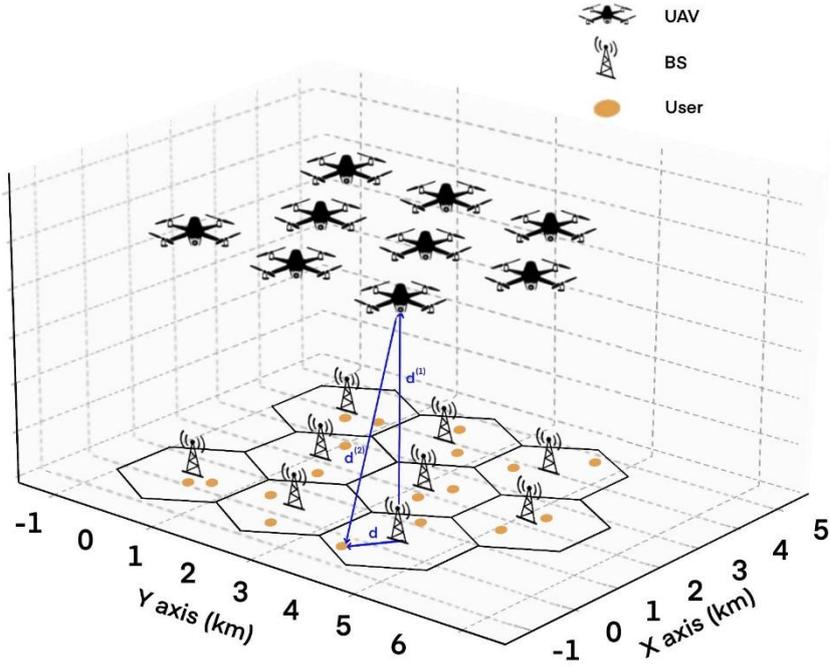

**Figure 1. System Model**

Within each cell, N active mobile users are uniformly distributed over the cell coverage area. The users are assumed to be randomly positioned at the beginning of each transmission frame and remain stationary during the frame duration. The distance $r$ between a randomly positioned user and its serving BS follows the probability density function (PDF) of uniform distribution within a hexagonal cell, given by:

$$f_R(r) = \frac{2r}{R^2}, \quad 0 \leq r \leq R \qquad (1)$$

The channel between any transmitter-receiver pair is characterized by both large-scale path loss and small-scale fading effects. The large-scale path loss $L(d)$ models the pathloss (large-scale fading) over distance $d$, while the small-scale fading coefficient $g$ captures rapid fluctuations due to multipath propagation.

For terrestrial BS-to-user links, we adopt the standard path loss model following 3GPP recommendations [3]. The instantaneous channel power gain incorporates small-scale fading effects modeled as Rayleigh distributed for rich scattering environments without dominant line-of-sight components. For air-to-ground UAV-to-user links, the high positioning of UAVs typically results in more favorable propagation conditions with a higher probability of line-of-sight transmission, leading to reduced path loss compared to terrestrial links at the same distances.

### 3.1. UAV-assisted communication

Unlike CCUs, CEUs generally suffer from poor channel conditions due to the longer distance from their serving BSs. To address this limitation, UAVs are deployed as aerial relay nodes to assist CEUs. In this configuration, $M$ UAVs are positioned at an altitude of $h$ above the BSs, with each UAV assigned to its nearest CEU to establish a one-to-one pairing. Any UAVs that remain unassigned are deactivated to avoid introducing unnecessary interference, thereby ensuring efficient resource utilization. For CCUs, the received signal originates solely from the direct BS-to-user transmission. This signal strength is affected by the BS transmit power, the channel fading, and the distance-dependent path loss. Since CCUs rely exclusively on their serving BSs, the received signal is modeled as:

$$S_c = P g_c L(d) \quad (2)$$

where P denotes the BS transmit power, $g_c$ is the small-scale fading coefficient, and $L(d)$ represents the path loss at distance $d$. This expression forms the baseline reference for evaluating CEUs and UAV-assisted scenarios.

In contrast, CEUs benefit from an additional transmission path provided by the UAV relay. The received signal at a CEU consists of two components: the direct BS-to-CEU signal and the UAV-assisted amplify-and-forward (AF) relayed signal. The BS first transmits a signal to the UAV, which is then forwarded to the CEU, resulting in a composite received power. This process can be expressed as:

$$S_e = P g_e L(d) + P g^{(1)} L(d^{(1)}) g^{(2)} L(d^{(2)}) \quad (3)$$

where the first term represents the direct BS-to-CEU link, and the second term accounts for the relayed BS-to-UAV and UAV-to-CEU link. This equation highlights the advantage of UAV relays in enhancing CEU performance.

Because spectrum resources are reused across BSs, every user experiences interference from neighbouring BSs. This inter-cell interference degrades the effective signal quality and becomes a dominant factor in determining system performance. It can be modelled as the aggregated signals received from all interfering BSs, expressed as:

$$I_b = \sum_{n=1}^{N-1} P_n g_n L(d_n) \quad (4)$$

where $P_n$ is the transmit power of interfering BS $n$, $g_n$ is the channel fading coefficient, and $L(d_n)$ is the corresponding path loss. This formulation captures the impact of co-channel interference across the cellular network.

Similarly, UAVs operating in the same frequency band also contribute additional interference to users. Each interfering UAV introduces a two-hop signal, consisting of the BS-to-UAV and UAV-to-user channels. The cumulative interference from these UAVs is mathematically expressed as:

$$I_u = \sum_{m=1}^{M-1} P \, g^{(m1)} L(d^{(m1)}) g^{(m2)} L(d^{(m2)}) \quad (5)$$

where $g^{(m1)}$ and $g^{(m2)}$ are the fading coefficients for the BS-to-UAV and UAV-to-user links of UAV $m$, respectively. This equation reflects the fact that UAV-assisted networks must carefully manage interference among aerial relays.

The overall interference experienced by a typical user is the combination of terrestrial and aerial interference. By aggregating the contributions from both BSs and UAVs, the total interference is expressed as:

$$I = I_b + I_u \qquad (6)$$

This compact representation serves as a key parameter in evaluating user performance.

With these definitions, the instantaneous downlink SINR of a CCU can be expressed. The SINR captures the effective quality of the received signal in the presence of interference and noise. For a CCU, it depends on the direct BS-to-user link quality, the interference from other BSs and UAVs, and the thermal noise power. The SINR is therefore given by:

$$SINR_c = \frac{P g_c L(d)}{I_b + I_u + \sigma^2} \qquad (7)$$

where $\sigma^2$ denotes the noise power. This expression serves as the benchmark case for evaluating CEUs.

For CEUs supported by UAV relays, the SINR includes both the direct and relayed signals in the numerator. This provides an additional diversity gain, which can significantly enhance link reliability and coverage. The denominator remains the same, since CEUs are affected by the same interference and noise conditions. The corresponding SINR is given by:

$$SINR_e = \frac{P g_e L(d) + P g^{(1)} L(d^{(1)}) g^{(2)} L(d^{(2)})}{I_b + I_u + \sigma^2} \qquad (8)$$

This formulation demonstrates the potential improvement of UAV assistance for CEUs.

Finally, the achievable downlink transmission rate for any user is directly related to its SINR. This relationship follows the Shannon capacity formula, which defines the maximum data rate achievable under given signal quality conditions. Hence, the user throughput for $i \in \{c, e\}$ is expressed as:

$$C_i = \log_2(1 + SINR_i) \qquad (9)$$

This equation provides the fundamental performance metric for comparing CCUs and CEUs under UAV-assisted communication.

### 3.2. Optimization Formulation

According to the 3GPP specifications, the transmission power of a base station (BS) is typically constrained within a predefined range, denoted as $(P_{min}, P_{max})$ limitation arises from both regulatory requirements and hardware capabilities, ensuring that the BS can provide sufficient coverage while avoiding excessive interference to neighboring cells. In this context, power control plays a crucial role in balancing spectral efficiency and interference management.

The primary goal of the optimization problem is to maximize the achievable sum-rate of the system while adhering to these transmission power constraints. To formalize this objective, the achievable rate of user $i$ is denoted as $C_i$, and the optimization problem is expressed as follows:

$$\max_{P} C_i \quad (10)$$

Here, the optimization is performed over the transmission power variable $P$, aiming to identify the power allocation that yields the maximum user rate.

However, the optimization is subject to a strict power constraint imposed on the BS. Specifically, the transmit power must remain within feasible limits to ensure compliance with practical deployment scenarios. This constraint can be represented mathematically as:

$$s.t. \quad (0 \leq P \leq P_{max}) \quad (11)$$

where $P_{max}$ presents the maximum allowable transmission power defined by the 3GPP standard.

## 4. DEEP Q-LEARNING FRAMEWORK

### 4.1. Markov Decision Process Formulation

We formulate the power control problem as a Markov Decision Process (MDP) defined by the tuple $(S, A, P, R, \gamma)$, where $S$ is the state space, $A$ is the action space, $P$ is the state transition probability, $R$ is the reward function, and $\gamma \in [0,1]$ is the discount factor. The MDP framework enables the DQN agent to learn optimal power allocation policies through sequential decision-making under uncertainty.

#### 4.1.1 State Space Design

The state at time slot $t$, denoted as $s_t \in S$, captures the instantaneous network conditions relevant for power control decisions. Specifically, the state vector consists of:

$$s_t = [d_1^t, d_2^t, \ldots, d_{NK}^t, g_1^t, g_2^t, \ldots, g_{NK}^t, u_1^t, u_2^t, \ldots, u_{NK}^t] \quad (12)$$

where:

- $d_i^t \in [0, R]$ represents the normalized distance between user $i$ and its serving BS at time $t$, scaled to $[0,1]$ by dividing by cell radius $R$.
- $g_i^t$ denotes the instantaneous channel gain (including both path loss and small-scale fading) for user $i$, normalized to $[0,1]$ based on maximum observable channel gain.
- $u_i^t \in \{0,1\}$ is a binary indicator where $u_i^t = 1$ if user $i$ is classified as a CEU (i.e., $d_i^t > D_0$) and receives UAV assistance, and $u_i^t = 0$ otherwise.

The total dimensionality of the state space is $3NK$, where $N = 9$ is the number of cells and $K = 2$ is the number of users per cell, resulting in a state vector of dimension 54. This compact representation captures the essential information needed for intelligent power control while maintaining computational tractability for the DQN.

#### 4.1.2 Action Space Definition

At each time slot, the DQN agent selects a transmission power level for each BS in the network. To ensure discrete optimization suitable for Q-learning, the continuous power range $[P_{min}, P_{max}] = [5,38]$ is discretized into $M = 10$ equally spaced levels:

$$A = \{P_1, P_2, \ldots, P_{10}\} \quad (13)$$

where $P_m = P_{min} + (m-1) \cdot \frac{P_{max} - P_{min}}{M-1}$ for $m = 1,2,\ldots,10$.

In our multi-cell scenario with $N = 9$ BSs, the global action space would be $A^9$, which contains $10^9$ possible joint power allocation configurations. To address this exponential complexity, we employ a centralized DQN agent that learns a policy mapping states to joint power allocations for all BSs simultaneously. The action at time $t$ is denoted as:

$$a_t = [P_1^t, P_2^t, \ldots, P_N^t] \in A^N \quad (14)$$

### 4.1.3 Reward Function

The reward function is designed to maximize the overall network spectral efficiency while accounting for both CCU and CEU performance. At time slot $t$, after executing action $a_t$, the immediate reward is defined as the sum spectral efficiency across all users:

$$R_t = \sum_{n=1}^{N} \sum_{k=1}^{K} C_{n,k}^t = \sum_{n=1}^{N} \sum_{k=1}^{K} \log_2(1 + \text{SINR}_{n,k}^t) \quad (15)$$

where $C_{n,k}^t$ is the achievable rate of user $k$ in cell $n$ at time $t$, and $SINR_{n,k}^t$ is computed according to equations (6) or (7) depending on whether the user is a CCU or CEU.

This reward formulation encourages the agent to find power allocations that maximize aggregate throughput while implicitly balancing the trade-off between CCU and CEU performance through the logarithmic utility function, which provides diminishing returns for high-SINR users and emphasizes improvements for low-SINR users.

## 4.2. Deep Q-Network Architecture

The Deep Q-Network approximates the optimal action-value function $Q^*(s, a)$ using a feed-forward neural network with parameters $\theta$. The Q-function represents the expected cumulative discounted reward when taking action $a$ in state $s$ and following the optimal policy thereafter:

$$Q^*(s, a) = E[s_t = s, a_t = a] \quad (16)$$

The network architecture is designed to handle the high-dimensional state space while maintaining computational efficiency:

- **Input layer**: Accepts the state vector $s_t \in R^{54}$ (dimension $3NK = 54$)
- **First hidden layer**: 128 neurons with ReLU activation function $f(x) = (0, x)$
- **Second hidden layer**: 64 neurons with ReLU activation
- **Output layer**: $|A^N| = 10^9$ neurons (in practice, we use action sampling techniques to reduce computational complexity)

The network outputs Q-values $Q(s, a; \theta)$ for all possible actions $a \in A^N$, representing the estimated long-term value of each power allocation configuration in the current state.

To stabilize training, we employ a separate target network with parameters $\theta^-$ that is periodically updated from the main network. This dual-network architecture reduces oscillations in the learned Q-function by providing stable target values during training.

## 4.3. Training Algorithm

The DQN agent is trained using experience replay and $\epsilon$-greedy exploration. The complete training procedure consists of the following steps:

### 4.3.1 Initialization

- Initialize main Q-network parameters $\theta$ randomly
- Initialize target Q-network parameters $\theta^- = \theta$
- Initialize replay buffer $D$ with capacity 50,000
- Set initial exploration rate $\epsilon = 1.0$
- Set discount factor $\gamma = 0.99$
- Set learning rate $\alpha = 0.001$ for Adam optimizer

### 4.3.2 Episode Execution

For each episode $e = 1, 2, \dots, E$:

**Step 1: Environment Reset**

- Randomly generate new positions for all $NK$ users within their respective cells.
- Initialize state $s_0$ based on user distances, channel gains, and UAV assignment indicators

**Step 2: Time Slot Iteration**

For each time slot $t = 0, 1, \dots, T-1$ (where $T = 50$ time slots per episode):

**Action Selection**: Choose action $a_t$ using $\epsilon$-greedy policy:

$$a_t = \begin{cases} random\ action\ from\ A^N & with\ probability\ \epsilon \\ \arg\max_a Q(s_t, a; \theta) & with\ probability\ 1 - \epsilon \end{cases} \quad (17)$$

**Action Execution**:

- Set BS transmission powers according to $a_t = [P_1^t, \dots, P_N^t]$
- Compute SINR for all users using equations (6)-(7)
- Calculate immediate reward $R_t$ using equation (12)

**State Transition**:

- Update small-scale fading coefficients $\{g_i^{t+1}\}$ by sampling new Rayleigh realizations
- Observe next state $s_{t+1}$

**Experience Storage**: Store transition $(s_t, a_t, R_t, s_{t+1})$ in replay buffer $D$

**Network Training**: If $t \mod 10 = 0$ (every 10 time slots):

- Sample mini-batch of 256 transitions $\{(s_j, a_j, R_j, s_{j+1})\}$ from $D$
- Compute target Q-values:

$$y_j = R_j + \gamma Q(s_{j+1}, a'; \theta^-) \quad (18)$$

- Update main network parameters by minimizing mean squared Bellman error:

$$L(\theta) = \frac{1}{256} \sum_{j=1}^{256} \left( y_j - Q(s_j, a_j; \theta) \right)^2 \quad (19)$$

using Adam optimizer with learning rate $\alpha = 0.001$

**Target Network Update:** If $t \mod 100 = 0$:

$$\theta^- \leftarrow \theta \quad (20)$$

**Step 3: Exploration Rate Decay**

$$\epsilon \leftarrow (0.01, \epsilon \cdot 0.995) \quad (21)$$

The training continues until the performance converges, typically requiring 500-1000 episodes. The trained Q-network is then deployed for testing, where actions are selected greedily ($\epsilon = 0$) to maximize expected performance.

## 5. PERFORMANCE EVALUATION

### 5.1. Simulation Setup

We consider a cellular network with $N = 9$ cells arranged in a regular grid layout. At the center of each cell, a BS serves $K = 2$ users that are uniformly and randomly distributed within a distance range from 0.01 km to 2 km. Additionally, we deploy 9 fixed-position UAVs at height $h = 0.1$ km to enhance coverage and support joint transmission with the BS, particularly for CEUs.

Channel modeling includes small-scale fading represented by Rayleigh fading and large-scale fading characterized by the standard path-loss model $37.6(d)$, where $d$ denotes the distance between transmitter and receiver in meters. The AWGN power at the receiver is set at $-114$ dBm. Transmission power is discretized into 10 levels, varying from a minimum power of 5 dBm to a maximum power of 38 dBm.

The implemented DQN employs a four-layer feed-forward neural network with hidden layers comprising 128 and 64 neurons, respectively, and ReLU activation functions. Each simulation scenario is repeated across multiple episodes, with each episode containing 50 time slots. At the beginning of each episode, user positions are randomly regenerated to ensure robust training. Experience replay utilizes mini-batches of size 256, sampled from a replay buffer with a capacity of 50,000 experiences every 10 time slots.

### 5.2. Performance Evaluation

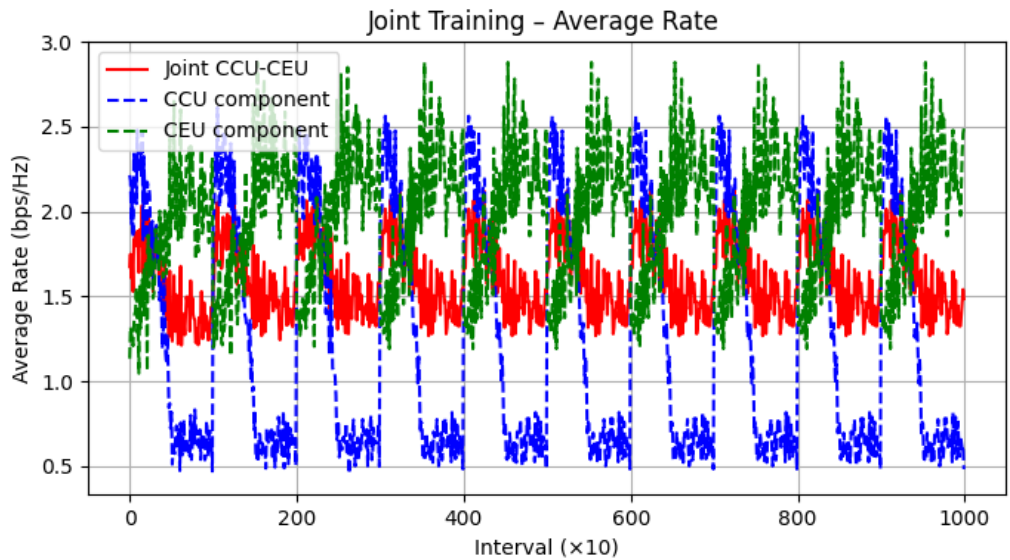

Figure 2. Training - Average rate vs time intervals

Figure 2 illustrates the training performance of the Deep Q-Learning based power control algorithm in terms of average rate for different user groups. The figure shows three curves: the CEU component, the CCU component, and the joint CCU–CEU users. It is evident that CEUs quickly achieve a stable performance level, converging to approximately 2.0–2.5 bps/Hz after several training episodes. This improvement comes from UAV assistance, which provides additional signal strength to users located far from the serving BSs. On the other hand, the CCU component experiences higher fluctuations and eventually stabilizes at around 0.5 bps/Hz. This degradation is mainly caused by the strong inter-cell and UAV interference affecting CCUs, even though they are closer to the BSs. As a result, the joint average rate of CCU and CEU users remains around 1.5 bps/Hz. These results highlight the trade-off introduced by UAV deployment: while CEUs benefit significantly, CCUs may experience reduced performance unless advanced interference management or optimized power allocation strategies are applied.

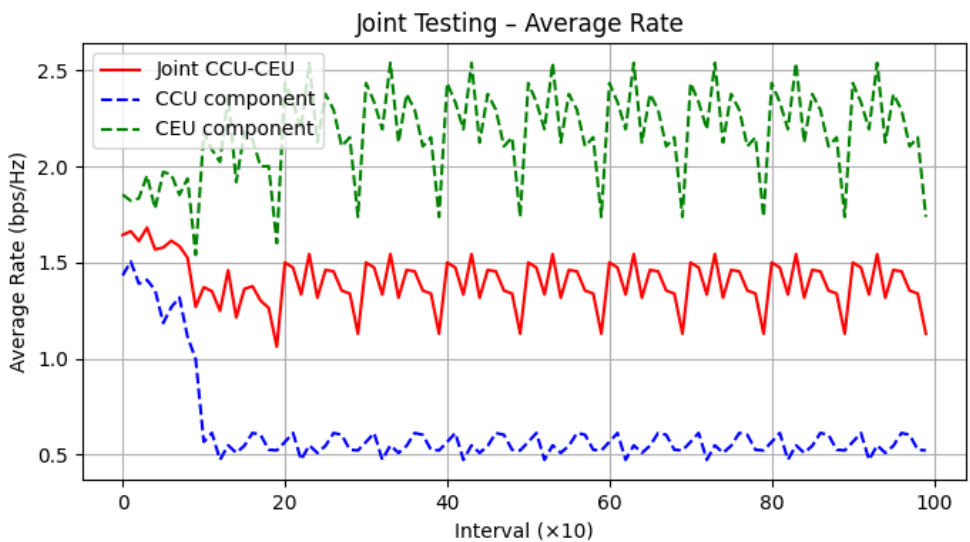

Figure 3. Testing - Average rate vs time intervals

Figure 3 presents the performance of the proposed UAV-assisted cellular network model during the testing phase after the Deep Q-Learning training process has been completed. The figure shows the evolution of the average transmission rate for three categories: the CEU component, the CCU component, and the joint CCU–CEU users.

From the results, it is evident that the CEU component achieves a stable average rate within the range of 2.0–2.5 bps/Hz. This demonstrates the effectiveness of UAV assistance in enhancing coverage for users located at the cell edge, where direct links to the BS usually suffer from severe path loss. With UAV relays providing additional signal paths, CEUs maintain consistently high throughput.

In contrast, the CCU component shows a noticeable degradation in performance compared to the training phase. Initially, the CCUs achieve rates close to 1.5 bps/Hz, but the average rate quickly drops and stabilizes at around 0.5 bps/Hz. This reduction is primarily caused by strong interference from neighboring BSs and UAV transmissions. Although CCUs are physically closer to their serving BSs and experience lower path loss, the presence of multiple interfering UAV signals significantly impacts their achievable rate.

As a result of this trade-off, the joint CCU–CEU average rate stabilizes at approximately 1.5 bps/Hz. While this indicates that UAVs substantially improve CEU performance, it also highlights a fairness issue, as CCU users experience reduced quality of service. These findings emphasize the importance of optimized power allocation and interference management strategies in UAV-assisted networks. Without careful coordination, the system risks improving CEU performance at the expense of CCUs, leading to unbalanced user experiences.

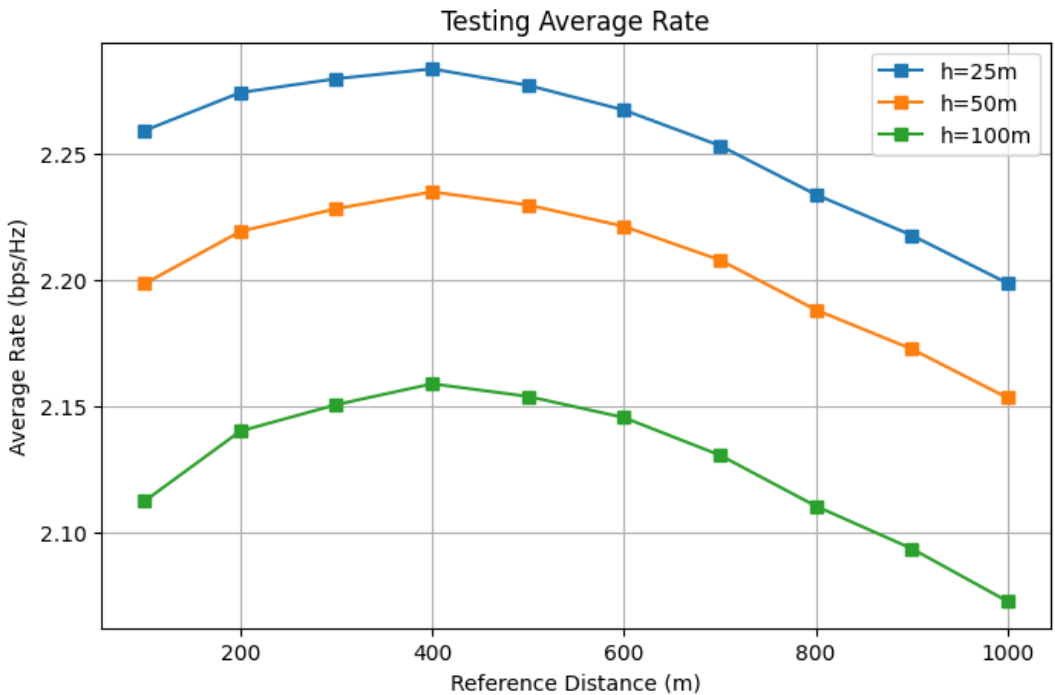

**Figure 4. Average rate vs reference distance**

Figure 4 illustrates the variation of the average rate with respect to the reference distance $D_0$ for UAV altitudes $h$ = 25 m, 50 m, and 100 m. At $D_0$ = 100 m, the average rates are approximately

2.26 bps/Hz ($h$ = 25 m), 2.20 bps/Hz ($h$ = 50 m), and 2.11 bps/Hz ($h$ = 100 m). In this region, nearly all users are classified as CEUs and receive UAV assistance. However, the high density of active UAVs introduces strong inter-cell interference, which limits the achievable performance despite the additional aerial support.

As the distance increases to $D_0$ = 300–400 m, the average rate reaches its peak: about 2.28 bps/Hz ($h$ = 25 m), 2.23 bps/Hz ($h$ = 50 m), and 2.16 bps/Hz ($h$ = 100 m). This indicates an optimal balance where only users who truly need assistance receive UAV support, while users closer to BSs rely on terrestrial links only. This selective deployment strategy maximizes the benefits of UAV assistance while minimizing unnecessary interference.

Beyond $D_0$=600 m, a gradual performance degradation is observed. At $D_0$ = 700 m, the average rate decreases to 2.25 bps/Hz ($h$ = 25 m), 2.21 bps/Hz ($h$ = 50 m), and 2.13 bps/Hz ($h$ = 100 m). This reduction becomes more significant at $D_0$ = 1000 m, where the average rate drops to 2.20 bps/Hz ($h$ = 25 m), 2.15 bps/Hz ($h$ = 50 m), and 2.07 bps/Hz ($h$ = 100 m). In this regime, nearly all users are classified as CCUs and do not receive UAV assistance, relying solely on terrestrial BS transmissions. As a result, users at the cell edge experience poor channel conditions without aerial support.

The comparison across altitudes further reveals that lower UAV height consistently provides higher average rate, with $h$ = 25 m outperforming $h$ = 50 m and $h$ = 100 m at every distance. This behavior can be explained by reduced path loss and stronger link quality at lower altitudes.

Comparing the proposed approach with benchmark schemes, the optimal configuration at $D_0$ = 400 m with $h$ = 25 m achieves 2.28 bps/Hz, which represents a 3.6% improvement over the no-UAV scenario ($D_0$ = 1000 m:2.20 bps/Hz) and a 0.9% improvement over the all-UAV scenario ($D_0$ = 100 m: 2.26 bps/Hz). These results demonstrate that selective UAV deployment based on distance threshold outperforms both extreme strategies, achieving better system performance through intelligent user classification and interference management.

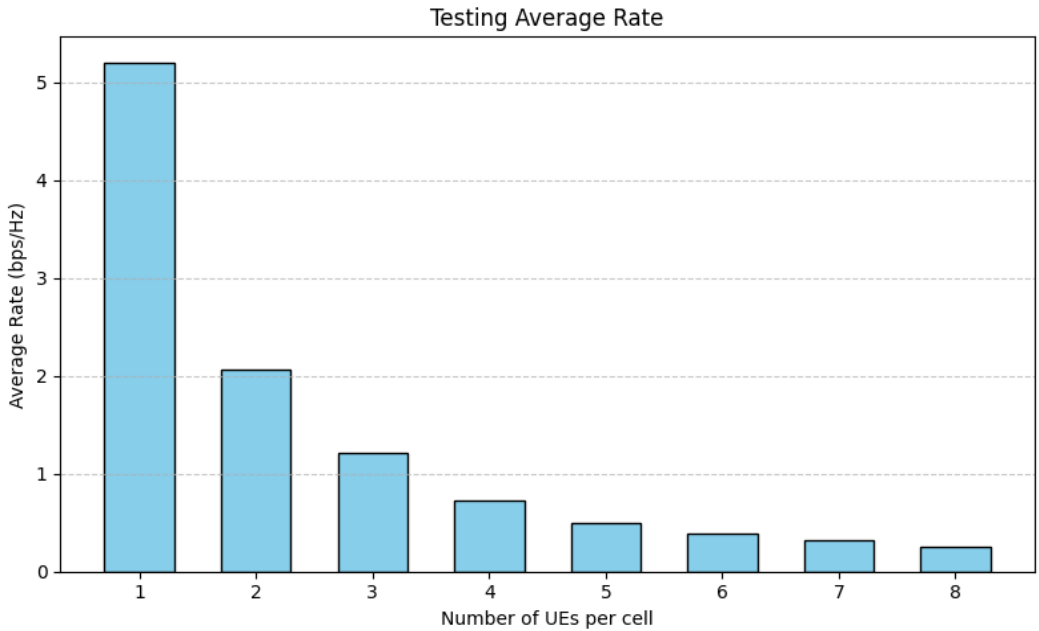

Figure 5. Average rate vs number of users

Figure 5 shows the average rate corresponding to the number of UEs per cell. It can be observed that as the number of UEs increases from 1 to 8, the average rate decreases significantly. When there is only one UE per cell, all resources are fully allocated to a single user, resulting in the highest achievable rate of approximately 5.2 bps/Hz. However, as the number of users increases, these resources must be shared among multiple UEs, leading to a reduction in the average throughput per user. At eight UEs per cell, the average rate drops to around 0.25 bps/Hz. In addition, the degradation is not only caused by resource sharing but also by the increase of interference among users and between adjacent cells.

## 6. CONCLUSION

This paper has proposed a UAV-assisted cellular network with intelligent power control based on selective user assistance. The key contribution is the distance-based deployment strategy where only users beyond a reference distance threshold receive UAV relay assistance, while users closer to base stations rely on terrestrial links only. This approach balances the benefits of UAV assistance with the costs of increased interference. To optimize power allocation across BSs, we have employed a Deep Q-Network learning framework that learns control policies without requiring accurate channel models. Simulation results demonstrate the effectiveness of the proposed approach. The system achieves a peak average rate of 2.28 bps/Hz at the optimal reference distance of 400 m with UAV altitude of 25 m. This represents a 3.6% improvement compared to conventional networks without UAV assistance and 0.9% improvement compared to networks where all users receive UAV support indiscriminately. The results show a clear non-monotonic relationship between reference distance and system performance: at very short distances (100 m), most users are CEUs with UAV support but suffer from high interference; at very long distances (1000 m), most users are CCUs without UAV support and rely solely on terrestrial links; the optimal balance occurs at intermediate distances (around 400 m) where UAVs assist only those users who truly need additional coverage.

Future research will focus on extending the current framework to more stochastic network environments, wherein the spatial distribution of base stations is modeled as a random process. Such an approach is expected to provide a more comprehensive understanding of network variability and its impact on overall system performance. Furthermore, integrating advanced reinforcement learning techniques such as Double Q-Learning or Soft Q-Learning will be considered to enhance learning efficiency and accelerate convergence toward optimal solutions.

## CONFLICTS OF INTEREST

The authors declare no conflict of interest.

## AUTHORS

Bach Hung Luu received the Bachelor of Electronic and Communication Engineering Technology and Master of Communication Engineering in 2022 and 2024, respectively from the University of Engineering and Technology (UET), Vietnam National University (VNU), Hanoi. His main interests are in stochastic geometry models for wireless communications, 5G, 6G and ISAC.

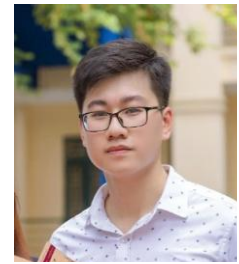

Sinh Cong Lam received the Bachelor of Electronics and Telecommunication (Honours) and Master of Electronic Engineering in 2010 and 2012, respectively, from the University of Engineering and Technology, Vietnam National University (UET, VNUH). He obtained his Ph.D. degree in Engineering from the University of Technology, Sydney, Australia. Dr. Lam Sinh Cong was appointed as an Associate Professor in Electronics in 2024. His research interests focus on modeling, performance analysis, and optimization for 5G and beyond 5G (B5G), as well as stochastic geometry models for wireless communications.

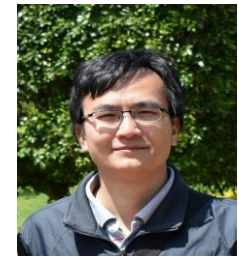

Nam Hoang Nguyen received the Ph.D. degree from the Vienna University of Technology, Austria, in 2002. He has been working with the University of Engineering and Technology, Vietnam National University, Hanoi, since 2011, where he was promoted to an Associate Professor in 2018. His research interests include resource management for mobile communications networks and future visible light communications.

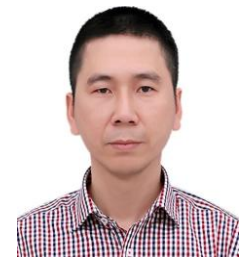